\documentclass{amsart}

\usepackage{graphicx}
\usepackage{amsmath}
\usepackage{amssymb}
\usepackage{color}

\usepackage{url}
\usepackage{graphicx}
\usepackage{multirow}

\theoremstyle{definition}

\theoremstyle{remark}

\numberwithin{equation}{section}

\begin{document}

%%
%% The title of the paper goes here.  Edit to your title.
%%

\title{Dynamics of Social Queues}

%%
%% Now edit the following to give your name and address:
%% 

\author{Hiroshi Toyizumi }
\address{Hiroshi Toyizumi, Waseda University, Japan}
\email{toyoizumi@waseda.jp}
%\urladdr{www.math.sc.edu/$\sim$howard} % Delete if not wanted.

%%
%% If there is another author uncomment and edit the following.
%%

\author{Jeremy Field}
\address{Jeremy Field, University of Sussex}
\email{j.field@sussex.ac.uk}
%\urladdr{www.math.sc.edu/$\sim$second}

%%
%% If there are three of more authors they are added in the obvious
%% way. 
%%

%%%
%%% The following is for the abstract.  The abstract is optional and
%%% if not used just delete, or comment out, the following.
%%%

\begin{abstract}
Queues formed by social wasps to inherit the dominant position in the nest are analyzed by using a transient quasi-birth-and-death (QBD) process.  We show that the extended nest life time due to division of labor between queen and helpers has a big impact for the nest productivity. 
\end{abstract}

%%
%%  LaTeX will not make the title for the paper unless told to do so.
%%  This is done by uncommenting the following.
%%

\maketitle

%%
%% LaTeX can automatically make a table of contents.  This is done by
%% uncommenting the following:
%%

%\tableofcontents

%%
%%  To enter text is easy.  Just type it.  A blank line starts a new
%%  paragraph. 
%%

%%
%% A new section is started as follows:
%%

\section{Introduction}
\label{sec:Introduction}
A wide variety of animals are known to form simple hierarchical 
groups called social queues, where individuals inherit resources or 
social status in a predictable order. Queues are often age-based, so 
that a new individual joins the end of the queue on reaching 
adulthood, and must wait for older individuals to die in order to 
reach the front of the queue. While waiting, an individual may work 
for her group, in the process often risking her own survival and 
hence her chance of inheritance. Eventually, she may survive to reach 
the head of the queue and becomes the dominant of the group.

Queueing has been particularly well-studied in hover wasps 
(Hymenoptera: Stenogastrinae) \cite{field2008ecology}. In hover wasp social groups, only one 
female lays eggs, and there is a strict, age-based queue to inherit 
the reproductive position.  While the dominant individual (queen) 
concentrates on breeding, subordinate helpers risk death by foraging 
outside the nest, but have a slim chance of eventually inheriting 
dominance.   Some explanations for this altruistic behavior and for 
the stability of social queues have been proposed and analyzed 
\cite{Field:2006fk,Kokko:1999qy}. Since both the productivity of the 
nest and the chance to inherit the dominant position depend 
critically on group size, queueing dynamics are crucial for 
understanding social queues, but detailed analysis is lacking. Here, 
using hover wasps as an example, we demonstrate that some basic queueing 
theory and non-homogeneous birth and death 
processes are useful for analyzing queueing dynamics and the 
population demographics of social queues.  Our work leads to better 
understanding of how environmental conditions and strategic 
decision-making by individuals interact to produce the observed group 
dynamics; and in turn, how group dynamics affects individual 
decision-making.

\section{Existing Models of Social Queues}
Various hypotheses have been proposed for the somewhat paradoxical evolution of helping behaviour, where an individual at least temporarily forfeits its own chance to reproduce and instead helps to rear another individual's offspring.  A general explanation is that helpers are nearly always rearing the offspring of a relative, so that copies of the helper's genes are propagated through helping \cite{hamilton1964genetical}. But since the relative's offspring rarely carry as large a proportion of the helper's genes as would the helper's own offspring, natural selection should favour helping only if helpers compensate by being more productive than they would be nesting alone \cite{queller1996origin}. 

There are different ways in which this could happen, some of which rely on the relatively short lifespans of adult wasps compared with the long development time of their progressively fed immature offspring \cite{Field:2015fk}. The extended parental care (EPC) implicit in progressive feeding means that a mothers often dies before her offsprings mature \cite{queller1994extended}.  For a potential helper, staying in the natal nest and rearing half-matured broods of a relative's offspring may be more productive than starting a new nest and rearing her own brood, because brood that are already part-matured are more likely to reach adulthood before the group as a whole fails (HS: Headstart hypothsis \cite{Queller:1989fk}). A subtlely different idea is that if a helper dies young, any dependent offspring that she has only part-reared can be brought to adulthood by the other individuals still remaining in the group, whereas for a female nesting independently, an early death means total brood failure (AFR: assured fitness return: \cite{nonacs2006transactional}). Another explanation is that if a helper has a chance to eventually inherit dominant status, it may be worth waiting without immediate fitness return if the expected reproductive success as dominant is large enough to outweigh the chance of death while waiting in the queue (DFR: delayed fitness return \cite{Kokko:1999qy,Hanna-Kokko:2001ys,shreeves2002gsa}).  Further discussions of validity of these explanations can be found in \cite{nonacs2006transactional,Shen:2011fk,Shen:2010kl,field2008ecology,queller1996origin,queller1994extended}.

Existing explanations try to understand social queues from the evolutionary perspective of rational individual decision making, using rather simple mathematical models.  Here, we analyze social queue from a different perspective, that of nest or population productivity and survival. As well as the above explanations for helping, we test the effect of a fifth general characteristic of sociality in insects: division of labour (DOL). In a social nest, the dominant can concentrate on laying eggs, not risking her life by foraging away from the nest, while her helpers forage.  Because of this division of labour, the queen has a considerably longer lifespan than her helpers. We investigate whether this will also increase the lifespan of the nest and the total number of reproductives dispersing from it.  Note that DOL is different with EPC, because, in DOL, the queen does not necessary expect the helpers to rear her offsprings after her death. 

We model the details of nest productivity in the following section by using a transient quasi-birth-and-death process, and compare nest productivity under the various models discussed above.

%\begin{table}
%\caption{Explanations of Social Queues}
%\begin{center}
%\begin{tabular}{|l|l|}
%\hline
%Model   & Explanations\\
%\hline
%\hline
%HS: Headstart hypothsis &Rearing related half-matured broods is easy.\\
%\hline
%AFR: Assured Fitness Return &Half-matured broods you reared will be maintained by other helpers.\\
%\hline
%DFR: Delayed Fitness Return & Wait for future dominance.\\
%\hline
%EPC: Extended Parental Care  & Parent life is too short to rear broods.\\
%%\hline
%%Transaction Skew & \cite{Ratnieks:1992kx}& Tug-of-war between dominant and subordinates.\\
%\hline
%DOL: Extended Nest Life Time & Longer nest life span, due to the division of labor, improves the productivity.\\
%\hline
%
%\end{tabular}
%\end{center}
%\label{Table:Explanations of Social Queues}
%\end{table}%
%\pagebreak

\section{Quasi-Birth-and-Death Process for Nest History of Social Queues}
\begin{figure}[htbp]
\begin{center}
\includegraphics[width=12cm,clip]{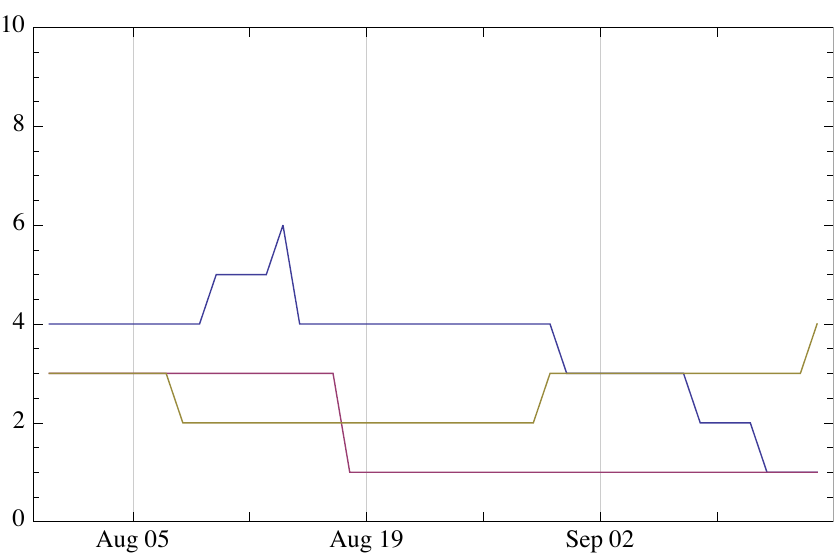}
\caption{Observed Dynamics of a Social Queue over a 6 week period in 2001; the blue, red and yellow lines represent the number of adults, pupae, and larvae respectively.}
\label{Fig:images/ExperimentRemoval/dyanamicsGraph.pdf}
\end{center}
\end{figure}

We use a transient quasi-birth-and-death (QBD) process to model not only by the number of adults but also the number of immature offspring (brood) on a nest. Figure \ref{Fig:images/ExperimentRemoval/dyanamicsGraph.pdf} shows an example of these dynamics in a real hover wasp nest.  QBD processes are intensively studied in the queueing literatures, especially in modelling complex communication systems (see \cite{LatoucheG:1999} for its good introduction).  By using QBD processes, we can keep track of the complex dynamics of populations such as social queue.  In QBD process, each event occurs at an exponentially-discrete time with its specific rate governed by the generator of the QBD process.

\begin{figure}[htbp]
\begin{center}
\includegraphics[width=13cm,clip]{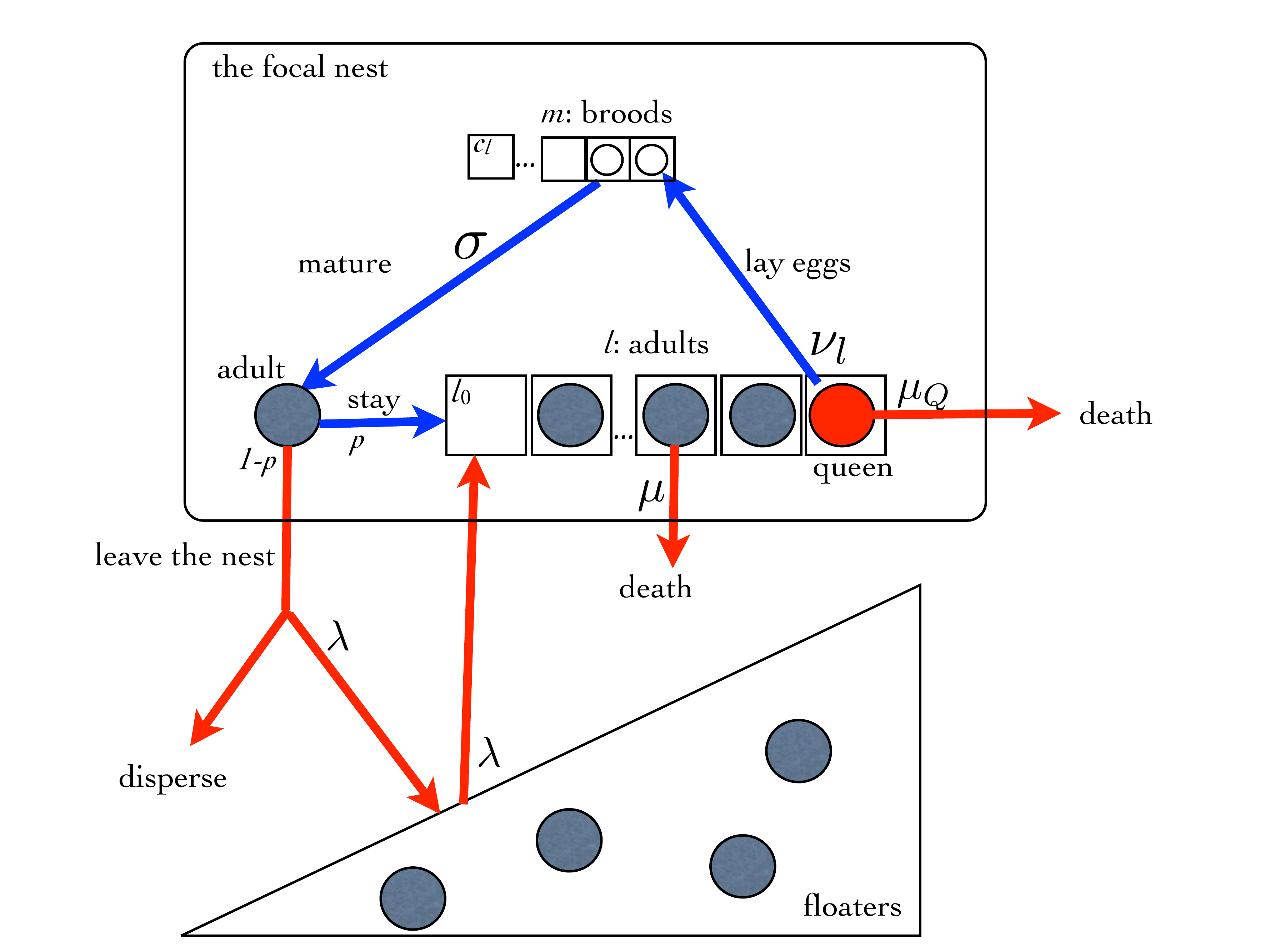}
\caption{Model of the Social Queue}
\label{images/QBDmodel.pdf}
\end{center}
\end{figure}

Considering a focal nest, we analyze its history and the productivity until the last individual dies and the nest is terminated (Fig. \ref{images/QBDmodel.pdf}).   We measure the nest productivity by the number of individuals that disperse from the nest and potentially initiate new nests. At time $0$, a founder individual builds the nest and starts the social queue.   Let $L(t)$ be the number of adults and $J(t)$ be the number of brood in the social queue at time $t$.   Note that we neglect males, which are not involved in nesting in wasps and bees.   The dominant queen, who is the most senior (oldest) adult in the nest, produces her brood one at a time with the rate $\nu_{l}$ where $l$ is the number of adults (including the queen herself) in the nest.  When instantaneous productivity is linear with queue size \cite{shreeves2002gsa}, $\nu_{l} = l \nu$.   For mathematical simplicity, we ignore the age of brood, and each $J(t)$ brood becomes adult with the rate $\sigma$ regardless of their age, which means that individual brood need an independent exponential time to become an adult.   Each adult forages and feeds brood.  In a nest with $l$ adults, at most $c_{l}$ brood can be accommodated.  When one of the adults dies, the number of brood may exceed the limit  $c_{l}$, in which case the surplus of brood will be abandoned.  We assume $c_{l}$ is an increasing function of $l$.

A new emerging adult has two options: (1) stays in the natal nest and become a helper with probability $p$ or (2) leaves the nest and disperse with the probability $1-p$.  We assume a maximum number of adults can stay in the nest, denoted by  $l_{0}$.  When the adult population reaches $l_{0}$, all subsequent emerging adults disperse until the adult population declines due to deaths.  A dispersed individual has another two choices: (2a) becomes a floater or (2b) founds a new nest somewhere else.  The floater population can be regarded as a reservoir shared among all nests in the site, and floaters join a focal nest in the site according to Poisson process with the rate $\lambda$.   To compensate for the influx at rate $\lambda$ from the floater population, emerging individual from the nest should join the floater population with the average rate $\lambda$, or the floater population will not be stable. 

Helpers that forage have a death rate $\mu$, but the dominant queen, who does not have to forage, has a different death rate $\mu_{Q}$.   At the time $\tau$ when the last adult on the nest dies, all the brood in the nest are abandoned and the nest is terminated.

The features of the social queue QBD model are summarized in Table \ref{Table;Features of QBD}.

\begin{table}[htdp]
\caption{Features of the QBD social queue model}
\begin{center}
\begin{tabular}{|l|c|l|}
\hline
Features & How?  & Remark\\
\hline
\hline
Large group size (2+) & $l_{0}\geq 2$& Upper Limit of Group Size\\
\hline
Distinction of adults and broods& $(L(t),J(t))$& No distinction among eggs, larvae and pupa.\\
\hline
Linearity of Reproduction & $\nu_{l}=l\nu$ & Lays eggs at a rate proportional to the group size. \\
\hline
Long maturation time & $1/\sigma$ &Exponential time, with no ageing effect.\\
\hline
Brood capacity & $c_{l}$ & Maximum number of brood allowed with $l$ adults.\\
\hline
Staying ratio & $p$ & Random decision. \\
\hline
Floaters & $\lambda$ &The rate of Poisson arrival.\\
\hline
Division of labour & $\mu$ and $\mu_{Q}$&Mortality difference between queens and helpers.\\
\hline
Progressive Brood & $\tau$ & Broods abandoned at the termination of nest.\\
\hline
\end{tabular}
\end{center}
\label{Table;Features of QBD}
\end{table}%

\begin{figure}[htbp] %  figure placement: here, top, bottom, or page
   \centering  
     \includegraphics[width=13cm]{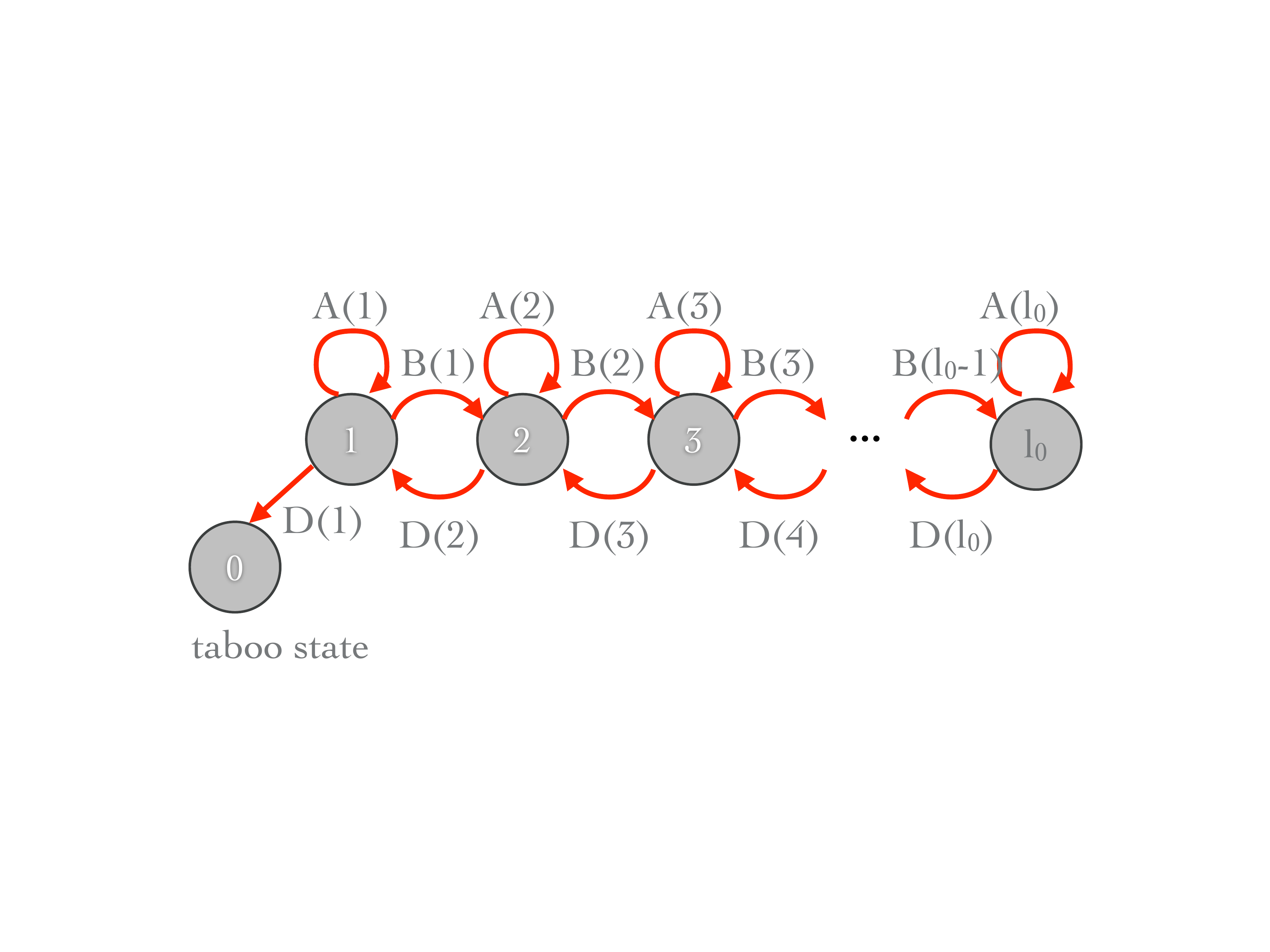}
  \caption{Transition Diagram of the matrix $\mathbf Q$.}
   \label{images/RateMatrixQBD.pdf}
\end{figure}

We assume $(L(t), J(t))$ forms a QBD process with the level $L(t)$ and its phase $J(t)$.   The process has $\sum_{l=1}^{l_{0}} (c_{l}+1)$ states;
\begin{align}
 \{
 \underbrace{(1,0),(1,1),(1,2), \dots, (1,c_{1})}_{l=1}, 
 \underbrace{(2,0),(2,1),\dots, (2,c_{2})}_{l=2},
 \dots,
 \underbrace{(l_{0},0),(l_{0},1),\dots, (l_{0},c_{l_{0}}}_{l=l_{0}})
 \}.
\end{align}
The termination time $\tau$ of the nest can be regarded as the hitting time to the boundary state $\{L(t) =0 \}$, and the social queue process $(L(t), J(t))$ is the taboo process.  Define the state probability of the social queue;
\begin{align}
p_{(l,j)}=p_{(l,j)}(t)= P\{ (L(t), J(t)) = (l,j), t \leq \tau \}.
\end{align}
We use the following convention to map the two-dimensional state probabilities $\{p_{(l,j)}(t)\}_{(l,j)}$ to the vector $\boldsymbol p(t)$;
\begin{align}
\boldsymbol p(t)&= (\boldsymbol p_{1}(t),\boldsymbol p_{2}(t),\dots, \boldsymbol p_{l_{0}}(t)) \notag \\ 
&= (\underbrace{p_{(1,0)},p_{(1,1)},p_{(1,2)}, \dots}_{l=1}, \underbrace{p_{(2,0)},p_{(2,1)},p_{(2,2)}, \dots}_{l=2}, \cdots, \underbrace{p_{(l_{0},0)},p_{(l_{0},1)},p_{(l_{0},2)}, \dots}_{l=l_{0}}).
\end{align}
The founder starts the nest at time $0$ and $(L(0), J(0)) = (1,0)$, so the initial probability vector is 
\begin{align}
\boldsymbol p(0) = (1,0,0, \dots, 0,0,0 \dots,  0,0,0).
\end{align}
The dynamics of social queue QBD processes are described by the following Kolmogorov equation:
\begin{align}\label{eq:Kolmogorov equation}
\frac{d}{dt} \boldsymbol p(t) = \boldsymbol p(t) \mathbf Q,
\end{align}
where $\mathbf Q$ is the infinitesimal generator of QBD process and defined by 
\begin{align}
\mathbf Q = \left(
\begin{array}{cccccc}
\mathbf A(1) & \mathbf B(1) & \mathbf 0 &  & \cdots  & \mathbf 0\\
\mathbf D(2) & \mathbf A(2)  & \mathbf B(2) & \mathbf 0 &  \cdots & \mathbf 0 \\
\mathbf 0&   \mathbf D(3) & \mathbf A(3)  & \mathbf B(3) &   \cdots &  \mathbf 0    \\
      &  &  \ddots  &  && \\
\mathbf 0 & \cdots & \mathbf D(l_{0}-2)& \mathbf A(l_{0}-2)  & \mathbf B(l_{0}-2) & \mathbf 0 \\
 \mathbf 0 &  \cdots &  \mathbf 0 & \mathbf D(l_{0}-1)& \mathbf A(l_{0}-1) & \mathbf B(l_{0}-1)  \\
 \mathbf 0 &  \cdots &  \mathbf 0  &\mathbf 0 & \mathbf D(l_{0})& \mathbf A(l_{0})
\end{array}
\right).
\end{align}
Here $\mathbf B(l)$, $\mathbf A(l)$, $\mathbf D(l)$ and $\mathbf 0$ are submatrices, and each represents a specific movement of social queue summarized in Table \ref{Table:Transition Rate Matrices} (see also Figure \ref{images/RateMatrixQBD.pdf}).  
\begin{table}[htdp]
\caption{Transition Rate Matrices.}
\begin{center}
\begin{tabular}{|c||l|l|}
\hline
matrix& transition & meaning \\
\hline
\hline
$\mathbf Q$ & $(l,j)\to (m,i) $&Generator of social queue QBD process. \\
\hline
$\mathbf B(l)$ &$(l,j)\to (l+1,i)$& Emerge of an adult or joining of a floater.  \\
\hline
$\mathbf A(l)$&$(l,j)\to (l,i)$ & Birth of a brood or dispersal of an adult.\\
\hline
$\mathbf D(l)$&$(l,j)\to (l-1,i)$ & Death of an adult.\\
\hline
$\mathbf 0$&$(l,j)\to (m,i) $ & No transitions\\
\hline
\end{tabular}
\end{center}
\label{Table:Transition Rate Matrices}
\end{table}%

Since the matrix $\mathbf Q$ has the two-layered coordination system, its element is expressed as $\mathbf Q_{\{ (l,j), (m,i)\}}$.
 The matrix $\mathbf B(l)$ is $(c_{l}+1) \times (c_{l+1}+1)$-submatrix and represents the increase of adult population (transition $l \to l+1$) and is defined by 
\begin{align}
\mathbf B(l) = 
\bordermatrix{
&0&1&2&      & \cdots&  c_{l} &c_{l}+1 &\cdots & c_{l+1}\cr
0& \lambda  & 0 & 0 &     & \cdots&  0 &0 &\cdots & 0\cr
1& p \sigma  & \lambda  & 0  &   & \cdots&  0 &0 &\cdots & 0\cr
2& 0 & 2 p \sigma  & \lambda &   & \cdots&  0 &0 &\cdots & 0\cr
\vdots& &  & &  \ddots   &  && \cr
c_{l}-1& 0 &   \cdots & 0 & (c_{l}-1)  p \sigma  & \lambda  & 0  &0 &\cdots & 0\cr
c_{l}& 0 &   \cdots &  0 & 0 &c_{l} p \sigma  & \lambda &0 &\cdots & 0
}
\end{align}
The diagonal element $\lambda$ is the inflow rate from the floater population following a Poisson process, and the term $jp \sigma$ in the lower off-diagonal is the maturation rate of adults (an immature brood becomes an adult) which decide to stay on the nest (with probability $p$), resulting in the number of brood on the nest decreasing by one ($j \to j-1$).
The matrix $\mathbf A(l)$ represents the dynamics of brood that mature to produce adults which then leave the nest (the transition inside the level $l$), and is defined by

\begin{align}
\mathbf A(l) = 
\left(
\begin{array}{ccccccc}
a_{l,0}& \nu_{l}& 0 & 0   & \cdots& 0 \\
 (1-p) \sigma  & a_{l,1}  & \nu_{l}  & 0  & \cdots& 0\\
 0 & 2 (1-p) \sigma  & a_{l,2} & \nu_{l}   & \cdots& 0 \\
        &  & &  \ddots   &  && \\
  0 &  \cdots & 0 & (c_{l}-1) (1-p) \sigma  & a_{l,c_{l}-1}  & \nu_{l}  \\
 0   &  \cdots & 0& 0 &c_{l} (1-p) \sigma  & a_{l,c_{l}} 
\end{array}
\right),
\end{align}
for $l=1,2, \dots , l_{0}-1$.  The diagonal terms represents aggregated outbound flow (transition $(l ,j)\to $  other states) and when $l=2, \dots , l_{0}-1$, 
\begin{align}\label{eq: alj}
a_{l,j} = 
\begin{cases}
 - \lambda-(l-1) \mu -\mu_{Q}- \nu_{l} - j\sigma  &\text{ for $j \not= c_{l}$, }\\
- \lambda-(l-1) \mu -\mu_{Q} - j\sigma  &\text{ for $j = c_{l}$, }\
\end{cases}
\end{align}
and when $l=1$ 
\begin{align}\label{eq: alj}
a_{1,j} = 
\begin{cases}
 - \lambda-\mu - \nu_{1} - j \sigma &\text{ for $j \not= c_{1}$, }\\
- \lambda-\mu  - j \sigma &\text{ for $j = c_{1}$. }
\end{cases}
\end{align}

The upper off-diagonal terms $\nu_{l}$ represents the rate of brood production when there are $l$ adults, resulting in an increase in the number of broods ($j \to j+1$).  The term $j(1-p) \sigma$, which is the lower off-diagonal of $\mathbf A(l)$, is the maturation (emergence) rate of adults that disperse, again resulting in a decrease in the number of brood ($j \to j-1$).    The matrices $\mathbf A(1)$ and $\mathbf A(l_{0})$ correspond to boundary (the brink of termination and the saturated nest), and they have slightly different elements from other $\mathbf A(l)$, because they represent the extremes (boundaries) of the maximum and minimum possible adult population.   In the case of $\mathbf A(1)$, only the diagonal elements are different and represented in \eqref{eq: alj}, since a lone dominant queen ($l=1$) has to forage for herself.   Note $ a_{1,j}$ includes the outflow $\mu$ to the taboo state $\{l =0\}$.  On the other hand, for $\mathbf A(l_{0})$, at the maximum adult group size,
\begin{align}
\mathbf A(l_{0}) = 
\left(
\begin{array}{ccccccc}
a_{l_{0},0}&  \nu_{ l_{0}}& 0 & 0 &  0  & \cdots& 0 \\
 \sigma  & a_{l_{0},1}  & \nu_{ l_{0}}  & 0 & 0 & \cdots& 0\\
 0 & 2 \sigma  & a_{l_{0},2} & \nu_{ l_{0}}  & 0 & \cdots& 0 \\
        &  & &  \ddots   &  && \\
  0 & 0 &  \cdots & 0 & (c_{l_{0}}-1)\sigma  & a_{l_{0},(c_{l_{0}}-1)}  & \nu_{ l_{0}} \\
 0 & 0  &  \cdots & 0& 0 & c_{l_{0}} \sigma  & a_{l_{0},c_{l_{0}}} 
\end{array}
\right),
\end{align}
where the diagonal terms;
\begin{align}
a_{l_{0},j} = -(l_{0}-1) \mu - \mu_{Q}-\nu_{ l_{0}} - j\sigma,
\end{align}
reflecting the fact that no further floaters can join, and the lower off-diagonal elements $j \sigma$ reflects the fact that all emerging adults must disperse.  The matrices $\mathbf D(l)$ represent deaths (the transition $l \to l-1$) for $l \geq 2$.  The death rate of dominant queen ($\mu_{Q}$) is less than the death rate of helpers and lone queens  ($\mu$): helpers and lone queens have the same death rate \cite{shreeves2002gsa,Field:2000fk}.  Thus, \begin{align}
\mathbf D(l) = 
\{(l - 1) \mu + \mu_{Q}\} \mathbf I_{l \to l-1}  \text{ for $l \geq 2$,}
\end{align}
where $\mathbf I_{l \to l-1} $ is defined by
\begin{align}
\mathbf I_{l \to l-1} = 
\bordermatrix{
&0&1&      \cdots&  c_{l-1} \cr
0& 1  & 0 &     \cdots&  0 \cr
1& 0 & 1  &    \cdots&  0 \cr
\vdots&&&   \ddots   & \cr
c_{l-1}& 0 &   \cdots &0  & 1\cr
c_{l-1}+1& 0 &   \cdots   &0  & 1\cr
\vdots& &  & &    \vdots \cr
c_{l}& 0 &   \cdots   &0  & 1\cr
}.
\end{align}
The equation \eqref{eq:Kolmogorov equation} can be solved formerly, and
\begin{align}
\boldsymbol p(t) = \boldsymbol p(0) \exp \{ \mathbf Q t \},
\end{align}
where the exponential should be interpreted as the matrix exponential and
\begin{align}
\exp \{ \mathbf A \} = \sum_{n=0}^{\infty}\frac{\mathbf A^{n}}{n !}.
\end{align}

First we estimate $E[\tau]$, the expected time to the termination of the nest.  Since $P\{ L(t) = l, J(t) =j, \tau \geq t | L(0) = m, J(0) = i\} = \left( \exp\{\mathbf Q t\}\right)_{\{ (m,i), (l,j)\}}$, we have
\begin{align}
P\{ \tau \geq t\} = \boldsymbol p(0) \exp \{ \mathbf Q t \}\boldsymbol 1,
\end{align}
where $\boldsymbol 1 = (1,1,\dots,1)$.  Since $\mathbf Q$ is a sub-stochastic matrix, it has its inverse, and we have
\begin{align}
E[\tau]=\int_{0}^{\infty}P\{ \tau \geq t\}dt = \int_{0}^{\infty} \boldsymbol p(0) \exp \{ \mathbf Q t \}\boldsymbol 1dt 
= \boldsymbol p(0) (- \mathbf Q)^{-1}\boldsymbol 1.
\end{align}
Here we used the relation of integral of matrix exponential and the inverse: $(- \mathbf Q)^{-1} = \int_{0}^{\infty}\exp \{ \mathbf Q t \}dt $.
Let $T(l,j)$ be the cumulated time spent in $(l,j)$ until the termination of the nest.  Similarly, we can calculate its mean $E[T(l,j)]$ as
\begin{align}
E[T(l,j)]=E[\int_{0}^{\tau} 1_{\{(L,J)=(l,j)\}}(t)dt]= \boldsymbol p(0) (- \mathbf Q)^{-1} \boldsymbol 1_{(l,j)},
\end{align}
where $1_{S}(t)$ is the indicator function of the set $S$ and $\boldsymbol 1_{(l,j)}$ is the vector whose elements are all $0$ but only $(l,j)$-element is $1$.  

Next, we estimate the productivity of the nest.  Let $H$ be the number of adults dispersing from the focal nest.   Since adults mature at the rate $\sigma J(t)$, and stay in their natal nest with probability $p$ until $L(t)$ reached $l_{0}$, the conditional dispersal rate given $(L(t), J(t))=(l,j)$ is defined by
\begin{align}
r(l,j) = 
\begin{cases}
\sigma j &\text{for $l =l_{0}$},\\
(1-p)\sigma j &\text{otherwise}.
\end{cases}
\end{align}
Using this, we define the dispersal rate vector $\boldsymbol r$ whose elements are of the form $r(l,j)$.  Since
\begin{align}
E\left[ r(L(t),J(t)) 1_{\{t \leq \tau \}}(t)\right] = \boldsymbol p(0) \exp \{ \mathbf Q t \}  \boldsymbol r, 
\end{align}
we have
\begin{align}\label{eq:E[H] def}
E[H] &= E\left[\int_{0}^{\tau} r(L(t),J(t)) dt \right] =  \int_{0}^{\infty}E\left[ r(L(t),J(t)) 1_{\{t \leq \tau \}}(t)\right] dt \notag\\
 &= \int_{0}^{\infty}\boldsymbol p(0)\exp \{ \mathbf Q t \} \boldsymbol r dt 
=\boldsymbol p(0)  (- \mathbf Q)^{-1}\boldsymbol r.
\end{align}
We estimate the success of the social queue by $r_{nest}$, which is the net rate of growth in the number of individuals that disperse and produce new nests.  Since some dispersed individuals join the floater population with the average rate $\lambda$ and one original nest is terminated during the interval $\tau$, we have
\begin{align}
r_{nest}=\frac{E[H]-1}{E[\tau]}-\lambda.
\end{align}
If $r_{nest}$ is positive, the number of nests increases, and the larger $r_{nest}$ the more rapidly the population grows.

Now we check a simple analytically tractable but interesting example, which will be the basis of our analysis.  When $l_{0}=1$, $c_{1}=\infty$, $\nu_{1}=\nu$, $\lambda =0$ and $p=0$, the system is lone breeder model with no brood capacity limit, no helpers and no floaters.  All emerging wasps will disperse.   In this case, the brood population process $J(t)$ turns out to be a transient $M/M/\infty$ queue with the arrival rate $\nu$ and the departure rate $\sigma$ starting from $J(0)=0$.   It is well-known that the marginal distribution of a transient $M/M/\infty$ queue is a Poisson distribution, and
\begin{align}
P\{J(t)=j\}=\frac{\rho(t)^{j}}{j!}e^{-\rho(t)},
\end{align}
where $E[J(t)]= \rho (t) = \nu (1-e^{-\sigma t})/\sigma$.  Because the nest termination time $\tau$ is simply the lone breeder's exponential lifetime with the death rate $\mu$ independent of $J(t)$, $E[H]$ can be calculated directly as
\begin{align*}
E[H] &=E\left[  \int_{0}^{\tau}\sigma J(t)  dt \right]\\
&=\sigma \int_{0}^{\infty}E\left[ J(t) 1_{\{t \leq \tau \}}(t)\right] dt \\
&=\sigma \int_{0}^{\infty}E[J(t)] P\{t \leq \tau \}dt\\
&=\int_{0}^{\infty}\nu (1-e^{-\sigma t}) e^{-\mu t}dt\\
&= \frac{\nu}{\mu}- \frac{\nu}{\sigma+\mu}.
\end{align*}
Since $E[\tau]=1/\mu$, the growth rate of nests in a population of lone breeders is
\begin{align}
r_{nest}=\nu\left(1 - \frac{\mu}{\mu+\sigma} \right) -\mu.
\end{align}
For example, let $\mu=1$, $\nu=2$ and $\sigma=1/2$.  On average, over the period of her unit life time, the lone breeder will produce $2$ brood.  At first sight, $r_{nest}$ should be $\nu - \mu =1$.  However, on average, it takes two units of  time for brood to mature, and those brood that fail to mature before the death of the lone breeder (the ratio $\mu/(\sigma+\mu)=2/3$) will be ``wasted''.  Thus the net growth rate is negative and $r_{nest}=-1/3$.   The population of lone breeders will not  survive in this environment and go extinct.   Note that when $\sigma \to \infty$ (no need for progressive feeding), we have a mass provisioning model \cite{Field:2015fk}, and its productivity is $r_{nest}=\nu - \mu$.

\section{Numerical Examples of Social Queue Models}
A population of lone breeders faces extinction because of the long maturation time of brood, as seen above.  Now we determine whether extinction is still inevitable under social queueing.  We set the maximum adults in the nest $l_{0} = 7$.   The results are summarized in Table \ref{Table:Social Queue Models with Brood Capacity $c_{l} = 7 l$} and Figure \ref{Fig:images/NestProductivityComparison.pdf} and \ref{Figmages/QBD/QBD_SimpleSocialQueue.pdf}.

\subsection{Simple Social Queue}
In the simple social queue, emerging helpers rear the brood of the queen even after her death, taking into account of the effect of EPC (extended parental care).  There is a positive impact on $r_{nest}$, which improves to $0.14659$ $(= - 0.18671-(-1/3))$, but this impact is limited and $r_{nest}$ is still negative.  This is because it takes a long time to rear adult helpers, compared with the lifetime of the initially lone queen.  A simple social queue cannot solve the problem of longer offspring maturation time.  

\subsection{Social Queue with Division of Labor}
\begin{figure}[htbp] %  figure placement: here, top, bottom, or page
   \centering  
  \includegraphics[width=10cm]{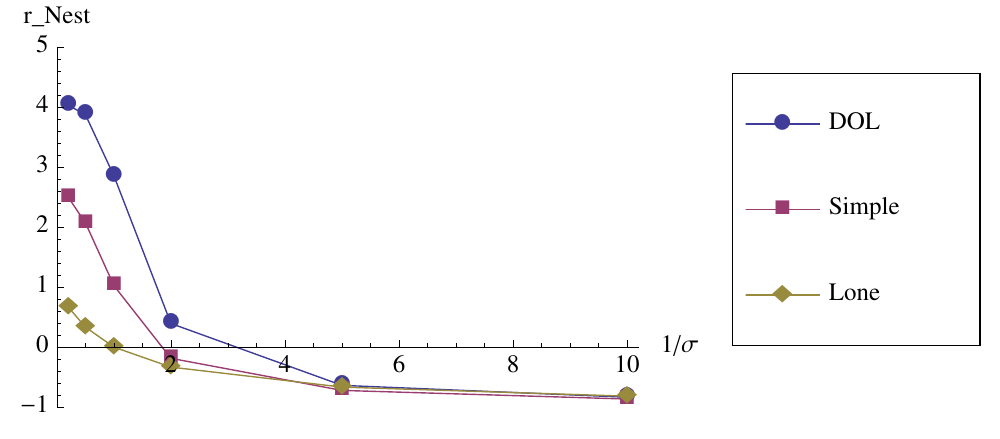}
  \caption{Comparison of the nest productivities for various maturation rates $\sigma$.  Lone, simple and DOL represent the nest productivity $r_{nest}$ of lone breeders, simple social queues and social queues with division of labor, respectively.  The other parameters are same as in Table \ref{Table:Social Queue Models with Brood Capacity $c_{l} = 7 l$}. }
   \label{Fig:images/NestProductivityComparison.pdf}
\end{figure}

\begin{figure}[htbp] %  figure placement: here, top, bottom, or page
   \centering  
   \includegraphics[width=6cm]{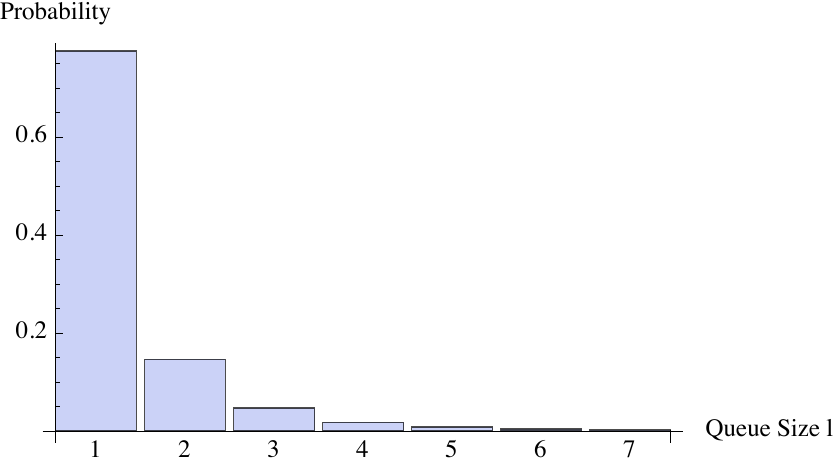}
   \includegraphics[width=6cm]{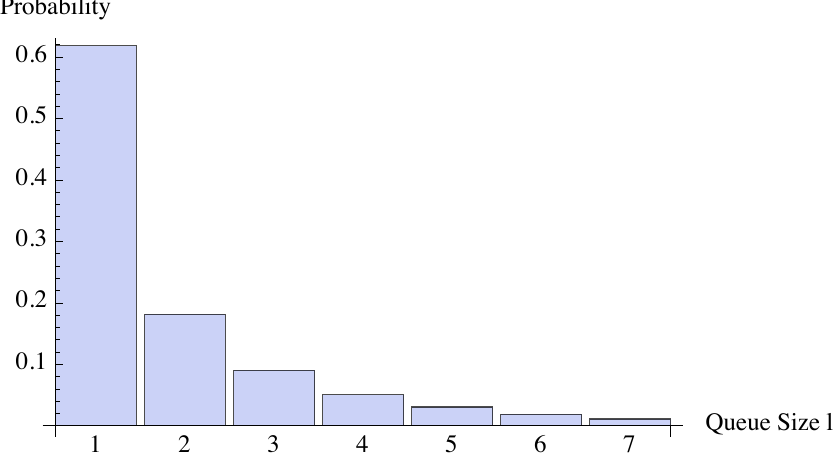}
   \includegraphics[width=6cm]{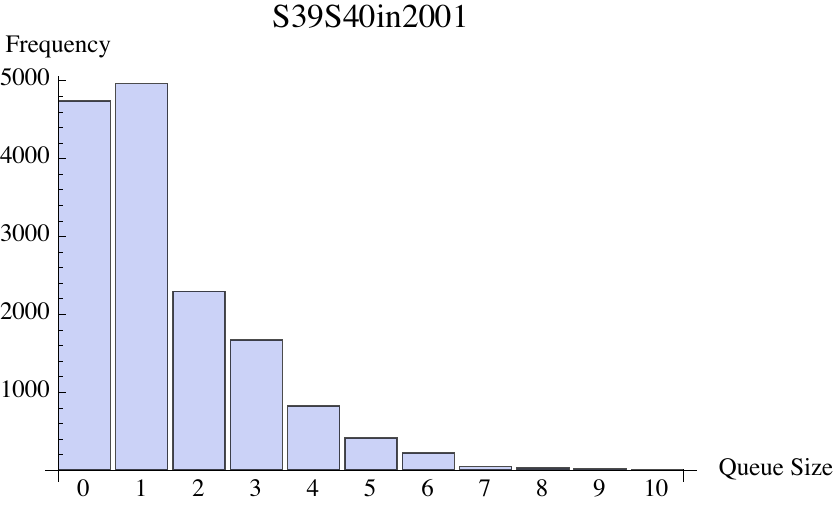}
  \caption{Distribution of the group size of adults in the nest. The first graph and second graph show the distribution derived by QBD model of simple social queue and social queue with DOL respectively.  The third graph shows the real distribution of group size observed during a 6 week period across 96 nests. }
   \label{Figmages/QBD/QBD_SimpleSocialQueue.pdf}
\end{figure}

A social queue with DOL (division of labor) as well as EPC can have positive $r_{nest}$.   Even though obtaining helpers is still rare, as seen in the second graph of Figure \ref{Figmages/QBD/QBD_SimpleSocialQueue.pdf}, once the queen gets helpers, the nest has longer time span and  the social queue can be productive.  The impact of the DOL is $0.580879 ( = 0.394169 - (-0.18671))$, which is considerably higher than the effect of EPC only.  See also Figure \ref{Fig:images/NestProductivityComparison.pdf} for comparing $r_{nest}$ for various maturation rates $\sigma$.  Note that lone breeders have negative $r_{nest}$ unless brood maturation time is shorter than expected adult life span.

\subsection{Effect of Floaters}
Floating seems strange behavior because they are wasting their breeding opportunity.  Floaters might be a backup for individual nests.  Thus, we checked the benefit of floating on $r_{nest}$.  As seen in Table \ref{Table:Social Queue Models with Brood Capacity $c_{l} = 7 l$}, floaters have an impact on improving $r_{nest}$ of especially for social queues adopting DOL.

\subsection{Staying Decision}
In the perspective of the nest productivity, all individual should stay in the nest until the nest to reach the full capacity $l_{0}$, where the nest has the maximum dispersal rate (see the column ``All Stay'' in Table \ref{Table:Social Queue Models with Brood Capacity $c_{l} = 7 l$}).  However, we can observe that almost half of emerging individual disperse even in a smaller queue size.  This might be explained by the balance of fitness benefit for individuals and nests.

\begin{table}[htdp]
\caption{Social Queue Models with Linear Brood Capacity $c_{l} = 5l$ and Linear Reproduction Rate $\nu_{l}=2l$. }
\begin{center}
\begin{tabular}{|l||c|c|c|c|c|c|c|c|c|c|c|c|c|}
\hline
Model&  $1/\sigma$& $p$ &$1/\mu$ &$1/\mu_{Q}$&$\lambda$& $E[\tau]$& $E[H]$ & $r_{nest}$\\
\hline
\hline
Lone Breeder ($c_{l}=\infty$) &2 & 0  & 1 & 1 & 0 &1 & 2/3 & -1/3 \\
\hline
Simple Social Queue & 2 & 1/2  & 1 & 1 & 0  &1.28801&0.759516& -0.18671\\
\hline
Social Queue with DOL  &  2 & 1/2  & 1 & 3 & 0 &1.61677& 1.63728& 0.394169 \\
\hline
with Floater (FLT) &  2 & 1/2  & 1 & 1& 1/10 & 1.40143& 0.961348& -0.12758\\
\hline
with DOL+ FLT &  2 & 1/2 & 1 & 3& 1/10 & 1.88688& 2.33406& 0.607017\\
\hline
All Stay  &  2 & 1 & 1 & 1& 0 & 4.26263&8.28762& 1.70965\\
\hline
\end{tabular}
\end{center}
\label{Table:Social Queue Models with Brood Capacity $c_{l} = 7 l$}
\end{table}%

%
%\begin{table}[htdp]
%\caption{Social Queue Models with Uniform Brood Capacity $c_{l} = 19$}
%\begin{center}
%\begin{tabular}{|l||c|c|c|c|c|c|c|c|c|c|c|c|c|}
%\hline
%Model& $\nu$ & $\sigma$& $p$ &$\mu$ &$\mu_{Q}$&$\lambda$& $E[\tau]$& $E[H]$ & $r_{nest}$\\
%\hline
%\hline
%Lone Breeder ($c_{l}=\infty$) & 2 & 1/2 & 0  & 1 & 1 & 0 &1 & 2/3 & -1/3 \\
%\hline
%Simple Social Queue & 2 & 1/2 & 1/2  & 1 & 1 & 0  &1.45427& 1.4478 &0.307923\\
%\hline
%Extended Life Time (ELT)  & 2 & 1/2 & 1/2  & 1 & 1/3 & 0 &2.44181 &5.23162&1.62934 \\
%\hline
%Floater (FLT)  & 2 & 1/2 & 1/2  & 1 & 1& 1/10 & 1.65158 & 2.01889 & 0.516921\\
%\hline
%ELT and FLT  & 2 & 1/2 & 1/2 & 1 & 1/3& 1/10 & 3.14118 & 7.89772 &2.0959\\
%\hline
%All Stay  & 2 & 1/2 & 1 & 1 & 1& 0 & 32.3325 & 86.7961 & 2.65356\\
%\hline
%All Female Stay  & 1 & 1/2 & 1 & 1 & 1/3 & 1/10 & 4.00806 & 2.09541 & 0.173303\\
%\hline
%\end{tabular}
%\end{center}
%\label{Table:Social Queue Models with Brood Capacity $c_{l} = 19$}
%\end{table}%

%\begin{acknowledgements}
%\end{acknowledgements}

\bibliography{/Users/toyo/Dropbox/References/economics,/Users/toyo/Dropbox/References/BioRef,/Users/toyo/Dropbox/References/virus,/Users/toyo/Dropbox/References/2011,/Users/toyo/Dropbox/References/2012,/Users/toyo/Dropbox/References/queue}

\begin{thebibliography}{10}
\expandafter\ifx\csname url\endcsname\relax
  \def\url#1{\texttt{#1}}\fi
\expandafter\ifx\csname urlprefix\endcsname\relax\def\urlprefix{URL }\fi

\bibitem{field2008ecology}
J.~Field, The ecology and evolution of helping in hover wasps (hymenoptera:
  Stenogastrinae), Ecology of Social Evolution (2008) 85--107.

\bibitem{Field:2015fk}
J.~Field, The evolution of progressive provisioning, Behavioral Ecology 16~(4)
  (July 2005) 770--778.
\newline\urlprefix\url{http://beheco.oxfordjournals.org/content/16/4/770.abstract}

\bibitem{Field:2006fk}
J.~Field, A.~Cronin, C.~Bridge, Future fitness and helping in social queues,
  Nature 441 (2006) 214--217.
\newline\urlprefix\url{http://dx.doi.org/10.1038/nature04560}

\bibitem{Field:2000fk}
J.~Field, G.~Shreeves, S.~Sumner, M.~Casiraghi, Insurance-based advantage to
  helpers in a tropical hover wasp, Nature 404~(6780) (2000) 869--871.
\newline\urlprefix\url{http://dx.doi.org/10.1038/35009097}

\bibitem{hamilton1964genetical}
W.~Hamilton, The genetical evolution of social behaviour. ii, Journal of
  theoretical biology 7~(1) (1964) 17--52.

\bibitem{Kokko:1999qy}
H.~Kokko, R.~A. Johnstone, Social queuing in animal societies: a dynamic model
  of reproductive skew, Proc. R. Soc. Lond. B 266 (1999) 571--578.
\newline\urlprefix\url{http://www.journals.royalsoc.ac.uk/openurl.asp?genre=article&id=doi:10.1098/rspb.1999.0674}

\bibitem{Hanna-Kokko:2001ys}
H.~Kokko, R.~A. Johnstone, T.~H. Clutton-Brock, The evolution of cooperative
  breeding through group augmentation, Proceedings of the Royal Society B:
  Biological Sciences 268~(1463) (2001) 187--196.
\newline\urlprefix\url{http://www.journals.royalsoc.ac.uk/content/en6x5dgrht76ulhw}

\bibitem{LatoucheG:1999}
G.~Latouche, V.~Ramaswami, Introduction to Matrix Analytic Methods in
  Stochastic Modeling, SIAM, 1999.

\bibitem{nonacs2006transactional}
P.~Nonacs, A.~Liebert, P.~Starks, {Transactional skew and assured fitness
  return models fail to predict patterns of cooperation in wasps}, American
  Naturalist 167~(4) (2006) 467--480.

\bibitem{queller1994extended}
D.~Queller, {Extended parental care and the origin of eusociality},
  Proceedings: Biological Sciences 256~(1346) (1994) 105--111.

\bibitem{queller1996origin}
D.~Queller, {The origin and maintenance of eusociality: the advantage of
  extended parental care}, in: Natural history and evolution of paper wasps.,
  Oxford University Press, Oxford, 1996, pp. 218--234.

\bibitem{Queller:1989fk}
D.~C. Queller, The evolution of eusociality: Reproductive head starts of
  workers, Proceedings of the National Academy of Sciences 86~(9) (1989)
  3224--3226.
\newline\urlprefix\url{http://www.pnas.org/content/86/9/3224.abstract}

\bibitem{Shen:2010kl}
S.-F. Shen, H.~Kern~Reeve, Reproductive skew theory unified: The general
  bordered tug-of-war model, Journal of Theoretical Biology 263~(1) (2010)
  1--12.
\newline\urlprefix\url{http://www.sciencedirect.com/science/article/B6WMD-4XRYTBM-3/2/15051acaa41548cf09231c01f66303a5}

\bibitem{Shen:2011fk}
S.-F. Shen, H.~Kern~Reeve, S.~L. Vehrencamp, Parental care, cost of
  reproduction and reproductive skew: A general costly young model, Journal of
  Theoretical Biology 284~(1) (2011) 24--31.
\newline\urlprefix\url{http://www.sciencedirect.com/science/article/pii/S0022519311002803}

\bibitem{shreeves2002gsa}
G.~Shreeves, J.~Field, {Group Size and Direct Fitness in Social Queues},
  American Naturalist 159~(1) (2002) 81--95.

\end{thebibliography}

\bibliographystyle{elsart-num-sort} 

\end{document}